\documentclass[%
 reprint,
 superscriptaddress,
 nobibnotes,
 amsmath,
 amssymb,
 aps,
 prb,
]{revtex4-1}

\usepackage{graphicx}
\usepackage{bm}

\usepackage{array}
    \newcolumntype{L}[1]{>{\raggedright\arraybackslash}p{#1}}
    \newcolumntype{R}[1]{>{\raggedleft\arraybackslash}p{#1}}
\usepackage{siunitx}
\AtBeginDocument{\usepackage{booktabs}}
\usepackage[flushleft]{threeparttable}    
\usepackage{color}

\AtBeginDocument{%
    \newwrite\bibnotes
    \def\bibnotesext{Notes.bib}
    \immediate\openout\bibnotes=\jobname\bibnotesext
    \immediate\write\bibnotes{@CONTROL{REVTEX41Control}}
    \immediate\write\bibnotes{@CONTROL{%
    apsrev41Control,author="08",editor="1",pages="1",title="0",year="1"}}
     \if@filesw
     \immediate\write\@auxout{\string\citation{apsrev41Control}}%
    \fi
}%

\newcommand{\bra}[1]{\left\langle #1 \right|}
\newcommand{\ket}[1]{\left|#1\right\rangle}

\begin{document}

\preprint{APS/123-QED}

\title{Charge qubit in van der Waals heterostructures}

\author{Bruno Lucatto}
 \email{brunolucatto@gmail.com}
 \affiliation{Grupo de Materiais Semicondutores e Nanotecnologia, Instituto Tecnol\'ogico de Aeron\'autica, DCTA, 12228-900 S\~ao Jos\'e dos Campos, Brazil}
\author{Daniel S. Koda} 
 \thanks{Current address: Department of Materials Science and Engineering, Massachusetts Institute of Technology, Cambridge, MA 02139}
 \affiliation{Grupo de Materiais Semicondutores e Nanotecnologia, Instituto Tecnol\'ogico de Aeron\'autica, DCTA, 12228-900 S\~ao Jos\'e dos Campos, Brazil}
\author{Friedhelm Bechstedt}
 \affiliation{Institut f\"ur Festk\"orpertheorie und -optik, Friedrich-Schiller-Universit\"at, Max-Wien-Platz 1, D-07743 Jena, Germany}
\author{Marcelo Marques} 
 \email{gmsn@ita.br} \affiliation{Grupo de Materiais Semicondutores e Nanotecnologia, Instituto Tecnol\'ogico de Aeron\'autica, DCTA, 12228-900 S\~ao Jos\'e dos Campos, Brazil}
\author{Lara K. Teles}
 \email{gmsn@ita.br}
 \affiliation{Grupo de Materiais Semicondutores e Nanotecnologia, Instituto Tecnol\'ogico de Aeron\'autica, DCTA, 12228-900 S\~ao Jos\'e dos Campos, Brazil}

\date{\today}

\begin{abstract}
\noindent

The use of spatial quantum superpositions of electron states in a gated vdW heterostructure as a charge qubit is presented.
We theoretically demonstrate the concept for the ZrSe$_2$/SnSe$_2$ vdW heterostructure using rigorous ab initio calculations.
In the proposed scheme, the quantum state is prepared by applying a vertical electric field, is manipulated by short field pulses, and is measured via electric currents.
The qubit is robust, operational at high temperature, and compatible with the current 2D technology.
The results open up new avenues for the field of physical implementation of qubits.

\end{abstract}

\maketitle


\section{Introduction}
The quantum superposition (QS) principle plays a major role in the so-called second generation of quantum technologies, which includes quantum counterparts of cryptography, imaging, computing, and sensing \cite{Georgescu2012}.
Preparation, manipulation, and measurement of the QS are central aspects to enable the operation of such advanced devices.
The superposition of two quantum states characterizes the unit of quantum information, a qubit, typically implemented by two-level systems\cite{Schneider2012}, polarization of light \cite{OBrien2003} or electron spin orientations\cite{Laucht2016}.

Solid-state quantum bits are attractive options such as spin qubits with electrically tunable spin-valley mixing in silicon\cite{Crippa2018,Bourdet2018}, the qubit of the two charge states of a negatively charged nitrogen vacancy in diamond\cite{Lucatto2017,Chou2018}, and the charge-qubit operation of an isolated double quantum dot\cite{Gorman2005}.
All these examples show that atomic-like structures in or of solids may be used as building blocks of future quantum computers or quantum information devices.
Recently, along with the strong development of two-dimensional (2D) material technology, progress has been made to find possible candidates for qubits in monolayer\cite{Pawowski2018} and multilayer\cite{Khorasani2016,Wang2019} structures.

One striking property of 2D materials is the formation of van der Waals (vdW) heterostructures, which consist of stacks of 2D crystals\cite{Geim2013,Novoselov2016,Liu2016}.
Despite the weak interaction between the two atomic layers in such a 2D heterostructure, if the band structures of the isolated sheets are nearly aligned on an absolute energy scale, QS can arise from wavefunctions localized in different layers but forming the conduction or valence bands of the heterostructure\cite{Koda2018}.
Bonding and antibonding combinations of orbitals localized on each subsystem build the basis functions of the joint conduction or valence band\cite{Koda2017}.
The resulting energy splittings and mixing coefficients of the wavefunctions depend on the vdW interlayer distance and the natural band discontinuities \cite{Koda2018}.
The mixing coefficients characterize the quantum-mechanical probabilities to find a certain carrier, electron or hole, in each of the two 2D materials of the heterostructure.
However, despite the fact that the QS property in 2D vdW heterostructures is highly promising, it has not yet been explored for qubit applications. 

In this Letter, we propose solid-state qubits based on Bloch states of a 2D vdW heterostructure.
In particular, we consider as a benchmark the system ZrSe$_2$/ SnSe$_2$ in which the qubit consists of a QS of the two first conduction states at M point.
These states can be manipulated by a vertical electric field oriented in the stacking direction, modifying the band alignments and, consequently, the mixing coefficients.
Thereby, the probabilities to measure a certain carrier on a specific side of the heterocombination can be modulated.
The time variation of the electric field allows for the control of the individual qubit state.
By means of \emph{ab initio} calculations, we demonstrate the effects of superposition as well as its manipulation via the gate field.
An example of a measurement scheme is presented.
The proposed system is robust and compatible with the current technology of 2D materials.

\begin{figure*}[t]
\centering
\includegraphics[width=\textwidth]{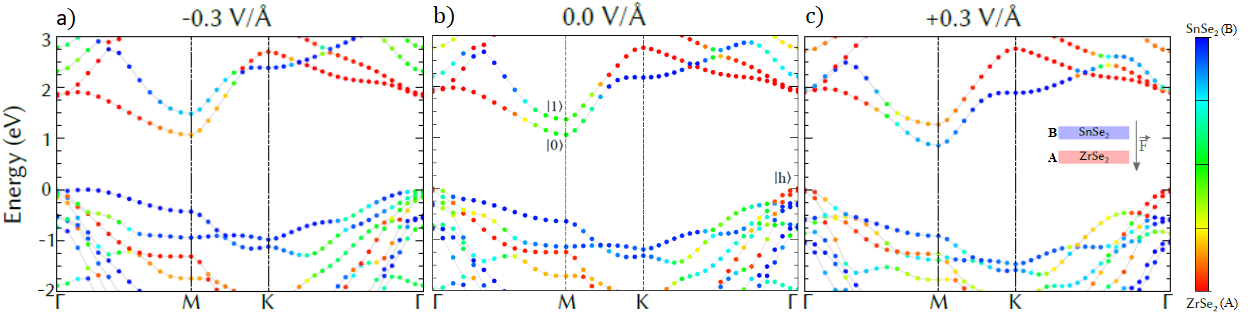}
\caption{Band structures of ZrSe$_2$/SnSe$_2$ heterostructure with applied vertical electric field of  (a) -0.3~V/\AA, (b) 0.0~V/\AA, and (c) +0.3~V/\AA. The inset in (c) shows the positive field orientation is considered from the SnSe$_2$ layer to the ZrSe$_2$ layer. The color of the marker represents the relative contribution of each monolayer to the eigenvalue. The VBM is chosen as energy zero. Band structures for intermediate electric fields are available in the Supporting Information.}
\label{bndstr}
\end{figure*}


\section{Effect of vertical electric field on band structure}

To illustrate the superposition of states for electrons and holes in biased 2D vdW heterostructures, we start with a model system consisting of ZrSe$_2$ and SnSe$_2$ transition metal dichalcogenide monolayers.
Because of the near lattice match, 1x1 cells with zero twist and small antisymmetric biaxial strain of $\pm$0.8\% are chosen \cite{Koda2017}.
A vertical electric field $\vec{F}$ simulates that the heterostructure is gated or vertically biased as displayed in the inset of Fig.~\ref{bndstr}c.
The band structures resulting for three field strengths are plotted in Fig.~\ref{bndstr} along with high-symmetry directions in the Brillouin zone (BZ) for a small energy interval around the fundamental gap.
The indirect semiconductor character with the conduction band minimum (CBM) at M and the valence band maximum (VBM) at $\Gamma$ is conserved for all field strengths.
In Fig.~\ref{bndstr}, the color of each eigenvalue represents the relative contribution of each crystal to the wavefunction.
It is obtained as the proportion of the projections of the Kohn-Sham orbitals onto the atomic orbitals (\emph{i.e.}, the orbital character of these levels\cite{Lucatto2017}) for all the atoms in each material.
In this work, the two-level system of the qubit is defined as the two lowest conduction states at the M point in the BZ.
The lowest conduction band state at M may be denoted by $\ket{0}$, whereas the next conduction band state $\ket{1}$ is higher in energy by ${\Delta_{ac}\approx0.3}$~eV but remains at the same $\vec{k}$ point.
The hybridization is also present in other states at the band edges, which may impact measures of carrier concentration.
In this sense, we also include in the following analysis the hole state in the VBM at $\Gamma$, denoted by $\ket{h}$.
These band states are composed by wavefunctions localized at one of the 2D crystals.
Figure~\ref{bndstr} clearly shows that their contribution can be manipulated by an external field $\vec{F}$.

\begin{figure*}[t]
\centering
\includegraphics[width=\textwidth]{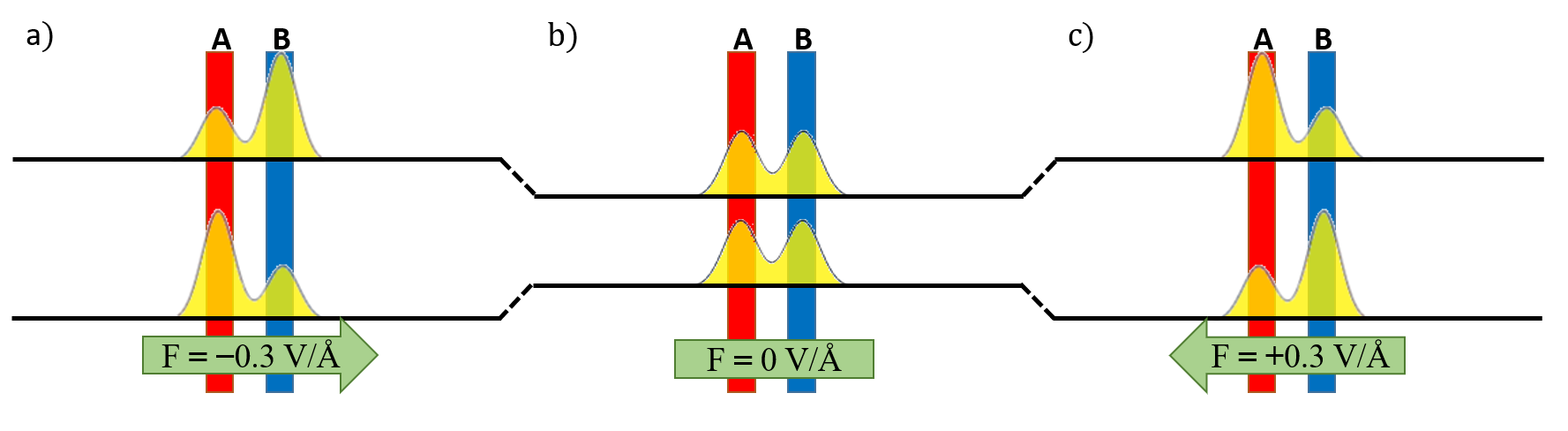}
\caption{Qualitative representation of the squared moduli of the wavefunctions for states $\ket{0}$ and $\ket{1}$ for (a) negative, (b) vanishing, and (c) positive electric fields.}
\label{wf}
\end{figure*}

In the unbiased case (Fig.~\ref{bndstr}b), $\ket{0}$ and $\ket{1}$ have almost equal contributions of each layer of the heterojunction as indicated by the green dots.
In the absence of an electric field, an electron in one of these states tends to be in an electronic state with equal probabilities to find the carrier in material A or B.
Instead, in the biased case, the conduction band states are given by superpositions of the wavefunctions that belong to material A or to material B with different weights as illustrated in Fig.~\ref{wf}.
This behavior is different from what happens at the top valence band state.
The character of $\ket{h}$ is predominantly from A, as indicated by the red color of the VBM in Fig.~\ref{bndstr}b.
Therefore, a hole tends to be localized in crystal A.

An external electric field in stacking direction of the vdW heterostructure can shift the bands of the crystals with respect to each other, as indicated in Figs.~\ref{bndstr}a and \ref{bndstr}c.
The corresponding change in energy also impacts the overlap of the orbitals, thus affecting their relative contribution to the bilayer wave function.
By applying an electric field in B-A direction, as indicated by the inset in Fig.~\ref{bndstr}c, electrons at $\ket{0}$ are lowered in energy.
The electric field then shifts down the band structure of material B relatively to the band structure of material A, consequently letting $\ket{0}$ be a state with a stronger character of material B, as indicated by Fig.~\ref{bndstr}c.
As $\ket{0}$ becomes more localized at B, $\ket{1}$ and $\ket{h}$ becomes more localized at A.
The consequences for the wave function localization are depicted in Fig.~\ref{wf}c.

The opposite holds true when we apply an electric field in the reverse direction: the bands of material B are shifted toward the vacuum level, allowing $\ket{0}$ to be more localized at A and $\ket{1}$ more localized at B, as indicated in Figs.~\ref{bndstr}a, and \ref{wf}a.
However, this is not the only effect observed in this case.
Since the band structure of material B is shifted toward higher absolute energies, the VBM of material B starts to line up with the VBM of material A, and thus $\ket{h}$ exhibits a stronger hybridization and nearly equal contributions from both sheets A and B, as indicated by the colors of Fig.~\ref{bndstr}a.
Moreover, for this extreme field situation, the VBM is slightly shifted from $\Gamma$, which is not true for the intermediate values (see the Supporting Information for a more detailed discussion on hole states).

\begin{figure}[b]
\centering
\includegraphics[width=0.45\textwidth]{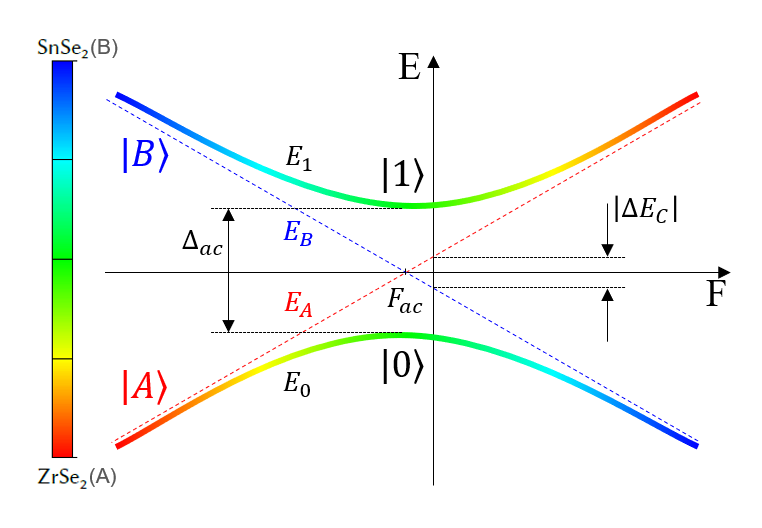}
\caption{Conduction-band energy level diagram versus gate-field strength $F$ formed by the localized electron states for an uncoupled system $\ket{A}$ and $\ket{B}$ (dashed lines) with eigenenergies $E_A$ and $E_B$, respectively. The hybridization of the states for the coupled system results in new eigenstates $\ket{0}$ and $\ket{1}$, with eigenenergies $E_0$ and $E_1$, respectively, and anticrossing energy $\Delta_{ac}$ (solid lines). For strong fields the qubit eigenstates are well approximated by $\ket{A}$ and $\ket{B}$, but for fields values around $F_{ac}$ the eigenstates are strongly delocalized. For the zero field, the energy difference $E_B-E_A=\Delta E_C$, characterizes the conduction band discontinuity.}
\label{Equbit}
\end{figure}

\begin{figure}[t]
\centering
\includegraphics[width=0.48\textwidth]{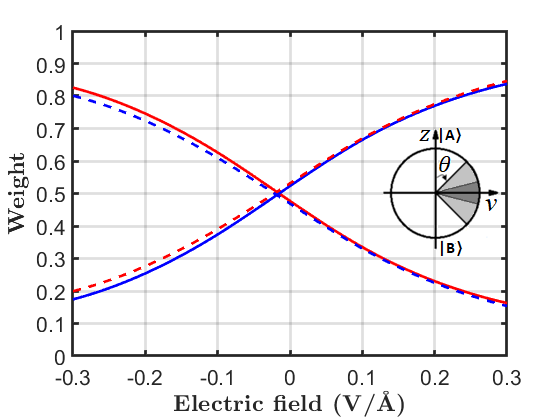}
\caption{Weights $|\alpha_\psi|^2$ (red) and $|\beta_\psi|^2$ (blue), for $\ket{\psi}$ equals $\ket{0}$ (solid lines) and $\ket{1}$ (dashed lines), as a function of the applied vertical electric field. The inset represents the area of the Bloch sphere in the A/B basis in which $\ket{0}$ and $\ket{1}$ are comprised for the considered electric fields (${|F|<0.3}$V/\AA) (light gray), and for small fields (${|F|<0.1}$V/\AA) (dark gray) considering a generic azimuthal angle~$\phi$. The horizontal axis indicates the direction of the vector ${\hat{v}=\cos\phi\hat{x}+\sin\phi\hat{y}}$ in the $xy$-plane.}
\label{character}
\end{figure}

There is a complementarity between $\ket{0}$ and $\ket{1}$ under the influence of the gate field, where one state becomes more localized in one sheet as the other becomes more localized in the other sheet.
Therefore, an electron occupying any superposition of the two states localized in different sheets configures a charge qubit in the AB heterostructure, where for strong positive electric fields the $\ket{0}$ and $\ket{1}$ states are localized in sheets B and A, respectively, and the opposite happens for strong negative electric fields.
The corresponding energy configuration is illustrated schematically in Fig.~\ref{Equbit} as a function of the electric field.
An anticrossing energy ${\Delta_{ac}=0.30}$~eV is determined by the difference between the eigenenergies $E_1$ and $E_0$ of the band states $\ket{1}$ and $\ket{0}$, respectively, for an electric field, where the layer contributions are equal.
This energy difference corresponds to a frequency $f\approx70$~THz.
In the studied system, this occurs at nearly vanishing field ${F_{ac}\approx-15}$mV/\AA, as depicted in Figs.~\ref{Equbit} and \ref{character}.
The conduction-level system, therefore, has similarities with the electronic structure of the charge qubit suggested in a double quantum dot \cite{Gorman2005}. 

The contribution of each layer A or B to the Bloch wave function of the states $\ket{0}$ and $\ket{1}$ in the AB heterostructure strongly depends on the electric field strength $F$.
Thus, an electron wave function $\ket{\psi(F)}$ of the AB heterostructure is mainly a combination of the corresponding wave functions $\ket{A}$ and $\ket{B}$ of the two individual atomic sheets with different weights.
It can be written as a superposition for a given field strength $F$
\begin{equation}
 \ket{\psi(F)}=\alpha_\psi(F)\ket{A}+\beta_\psi(F)\ket{B}
 \label{superposition}
\end{equation}
with complex coefficients. 
Because of the relatively large distance between the sheets, we consider the overlap of these states to be small, such that the normalization of the coefficients is given by ${|\alpha_\psi(F)|^2+|\beta_\psi(F)|^2=1}$.
Under this approximation, $\ket{A}$ and $\ket{B}$ are orthogonal and, therefore, the squared moduli of their coefficients give the weights of each sheet to the wavefunction, as illustrated in Fig.~\ref{character}.
This figure also provides evidence for the complementarity between $\ket{0}$ and $\ket{1}$, which further justifies the usage of the system as a charge qubit realized in the sheet arrangement.

The representation (\ref{superposition}) can be also interpreted as a coherent superposition of basic quantum states $\ket{A}$ and $\ket{B}$ at a given time or field strength, where the probability amplitudes $\alpha_\psi$, $\beta_\psi$ to find an electron characterize a linear combination as in a single qubit \cite{Schumacher1995}.
In a linear approximation around the state of maximum delocalization, considering $F_{ac}\approx0$, one finds for the biased ZrSe$_2$/SnSe$_2$ heterostructure 
\begin{equation}
  \left\{
    \begin{array}{l}
      |\alpha_0(F)|^2\\
      |\beta_0(F)|^2
    \end{array}
  \right\}
  =
  \left\{
    \begin{array}{l}
      |\beta_1(F)|^2\\
      |\alpha_1(F)|^2
    \end{array}
  \right\}
  = \frac{1\mp2.9F}{2}
  \text{ ($F$ in V/\AA).}
  \label{coefficients}
\end{equation}

The states $\ket{\psi(F)}$ of Eq.~(\ref{superposition}) can be described as a Bloch vector in the standard Bloch sphere representation, where the mixing coefficients are described by spherical coordinates with angles $\theta$ and $\phi$ as
\begin{align}
 \nonumber \alpha &= \cos(\theta/2)\\
 \beta &= e^{i\phi}\sin(\theta/2).
\end{align}
The polar angle $\theta$ can be calculated as
\begin{equation}
 \theta = 2\arccos(|\alpha|)=2\arcsin(|\beta|).
\end{equation}

For the considered values of the electric field, the Bloch vector lies in the shaded area depicted in the inset of Fig.~\ref{character}, which corresponds to the interval between $\theta\approx45^\text{o}$ and $\theta\approx135^\text{o}$.
Considering $|F|<0.1$V/\AA, the linear approximation introduces a small error, and the vector lies in the dark gray area in the inset of Fig.~\ref{character}.

\section{Usage as qubit}
A general superposition state $\ket{\psi}$ on the described two-level system for a given field $F$ can be expanded in the energy eigenvector basis $\{\ket{0},\ket{1}\}$ as ${\ket{\psi}=\xi\ket{0}+\eta\ket{1}}$.
Writing the energy eigenvectors in the A/B basis, ${\ket{0}=\alpha_0\ket{A}+\beta_0\ket{B}}$ and ${\ket{1}=\alpha_1\ket{A}+\beta_1\ket{B}}$, we find that $\ket{\psi}$ in this basis is written as ${\ket{\psi}=\alpha_\psi\ket{A}+\beta_\psi\ket{B}}$, where
\begin{align}
 \nonumber \alpha_\psi &= \xi\alpha_0+\eta\alpha_1\\
 \beta_\psi &= \xi\beta_0+\eta\beta_1.
\end{align}

A possible application of the two-level system as a quantum bit is to initialize the system in the desired state, by choosing a suitable electric field strength and allowing the system to relax to ensure the electron is in the lowest CBM, \emph{i.e.}, in state $\ket{0}$.
In order to apply single-qubit gate operations, the gate field is set to another value, thus changing the two-level system's Hamiltonian itself, since it is a function of the field strength $F$.
If this change is made within a slow process, the coefficients of the energy eigenvectors basis $\xi$ and $\eta$ would stay constant throughout the process, under the conditions of the adiabatic theorem.
Therefore, the electron would stay in state $\ket{0}$, regardless of the wavefunction of $\ket{0}$ being different from the starting one, which implies that $\alpha_\psi$ and $\beta_\psi$ are changed.
On the other hand, if the variation of the electric field is fast enough, the electron wavefunction would be approximately unchanged during the whole process, \emph{i.e.}, $\alpha_\psi$ and $\beta_\psi$ would be constant.
Together with the fact that the coefficients of the energy eigenstates in the A/B basis change with the electric field, this implies that $\xi$ and $\eta$ change.
This behavior opens the possibility of moving the electron to the excited state without recurring to optical excitations.
In order to measure the resulting state, the carrier concentrations in each sheet must be measured in a time window and compared to each other.

\begin{figure*}[t]
\centering
\includegraphics[width=0.75\textwidth]{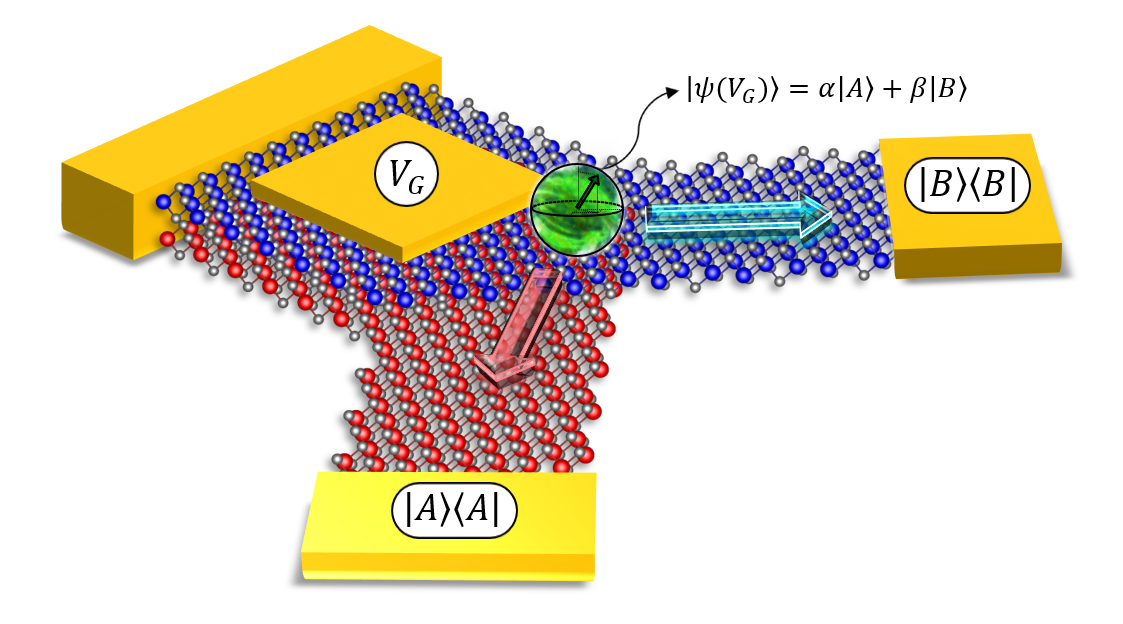}
\caption{Schematic representation of a possible physical implementation of the qubit. The electrode $V_G$ applies the field on the stacking direction, changing the state $\ket{\psi(V_G)}$ of the electron in the conduction band, represented by the green Bloch sphere. By inducing a drift in the horizontal direction, the electron wavefunction will collapse in one of the two electrodes of the isolated portion of the sheets ($\ket{A}\bra{A}$ or $\ket{B}\bra{B}$), with the probability depending on its localization in each sheet. The Zr, Sn and Se atoms are represented in red, blue and gray, respectively.}
\label{dev}
\end{figure*}

An example of operation would be to setup the system with a strong negative field, \emph{i.e.}, to start with ${\ket{0} \approx \ket{A}}$ and ${\ket{1} \approx \ket{B}}$, then suddenly change the electric field to a strong positive one.
Presuming the field switch happens fast enough, the electron state remains the same, \emph{i.e.}, $\ket{A}$ or $\ket{B}$.
However, after the switching, these site states correspond to the opposite energy eigenstates of the Hamiltonian for the new field strength, \emph{i.e.}, ${\ket{A} \approx \ket{1}}$ and ${\ket{B} \approx \ket{0}}$.
Therefore, by doing so, a Pauli-X quantum gate is applied to the qubit.
By applying an additional external bias in the horizontal direction in each sheet, a small carrier drift can be induced, and by measuring the resulting currents the carrier concentration in the layers can be deduced.
A possible structure for the device that would allow for the operations here described is depicted in Fig.~\ref{dev}.

Since the variation of the orbital character of a band is continuous with respect to the crystal momentum (see Fig.~\ref{bndstr}), even if more than one electron is excited, we can assume that it will have approximately the same mixing coefficients as the first one.
The Pauli exclusion principle is satisfied due to the difference in the crystal momentum quantum number.
Therefore, it may be possible to perform the same operation with many electrons at the same time, if the decoherence time does not decrease too much due to carrier collisions.
This would allow for single measurement operation, since the desired statistics of the results would be given by the relative amplitude between the currents flowing through each sheet.
Experimental realization of the qubit should provide a measure of how the (electro)chemical potential position affects the decoherence time.
Besides, since ${\Delta_{ac}>>k_BT}$, where $k_B$ is the Boltzmann constant and $T$ is the room temperature, we expect the system to operate at room temperatures.

It is worth emphasizing that the described vdW charge qubit is not restricted to the studied system.
Instead, due to the fact that there are several possible combinations of 2D materials, it is highly probable that other similar systems exist.
Considering that there is a practical limit to the intensity of electric fields that can be applied, which also limits how much bands can be shifted with respect to each other, a first filter to predict vdW combinations that may present QS is by analyzing their natural band alignments, as shown in Tables S1 and S2 of the Supporting Information.
However, for a real prediction, one needs to go further and make electronic structure calculations, since the desired hybridization effect depends not only on the proximity of the energy levels \cite{Koda2018}.
Moreover, applying strain on the heterostructure may also lead to hybridization in systems with large natural band discontinuities \cite{Koda2017}.

\section{Computational Details}
The structural and electronic properties are calculated using the density functional theory (DFT) as implemented in the Vienna Ab-initio Simulation Package (VASP) \cite{Kresse1996a}.
The wave functions and pseudopotentials are generated within the projector-augmented wave (PAW) method \cite{Kresse1999}.
Exchange and correlation (XC) are described using the Perdew-Burke-Ernzerhof (PBE) functional within the generalized gradient approximation (GGA) \cite{Perdew1996}.
Van der Waals interaction is taken into account using the optB86b functional \cite{klime2011}.
The repeated slab method is applied to simulate individual 2D crystals as well as their heterocombinations \cite{Bechstedt2003}.
Minimum lateral unit cells employed are found within the coincidence lattice method \cite{Koda2016}.
To account for the excitation aspect we add approximate quasiparticle corrections to the Kohn-Sham bands by applying the XC hybrid functional HSE06 \cite{Paier2006,Heyd2003,Heyd2006}.
More detailed information is given in the Supporting Information.

\section{Conclusion}
In summary, we identified the existence of spatial quantum superposition states in the conduction bands of van der Waals heterostructures with small natural band discontinuities and proposed their gate-field manipulation that can be employed in device applications.
Explicitly, we performed rigorous \emph{ab initio} calculations for the model vdW heterostructure consisting of atomic sheets of ZrSe$_2$ and SnSe$_2$ on which a variable vertical electric field was applied.
We obtained quantitative conduction band structures which demonstrate the feasibility of controlling the probability of the electron being on a specific side of the heterostructure by the external field.
Finally, we proposed to use the system as a charge qubit, which is based in a robust electronic state, does not require cryogenic operating temperatures, and is compatible with the preparation technology of 2D electronic devices.

\begin{acknowledgments}
The authors thank Dr. Ivan Guilhon for the fruitful discussions and revision of the text.
This work was funded by the Brazilian agencies FAPESP (grant n. 2012/50738-3), CAPES (PVE - grant n. 88887.116535/2016-00), and CNPq (grants n. 305405/2014-4, 308742/2016-8, and 154636/2016-9).
We acknowledge the National Laboratory for Scientific Computing (LNCC/MCTI, Brazil) for providing HPC resources of the SDumont supercomputer.
\end{acknowledgments}


\bibliography{references}

\end{document}


\preprint{APS/123-QED}

\title{Charge qubit in van der Waals Heterostructures - Supporting Information}

\author{Bruno Lucatto}
 \email{brunolucatto@gmail.com}
 \affiliation{Grupo de Materiais Semicondutores e Nanotecnologia, Instituto Tecnol\'ogico de Aeron\'autica, DCTA, 12228-900 S\~ao Jos\'e dos Campos, Brazil}
\author{Daniel S. Koda} 
 \thanks{Current address: Department of Materials Science and Engineering, Massachusetts Institute of Technology, Cambridge, MA 02139}
 \affiliation{Grupo de Materiais Semicondutores e Nanotecnologia, Instituto Tecnol\'ogico de Aeron\'autica, DCTA, 12228-900 S\~ao Jos\'e dos Campos, Brazil}
\author{Friedhelm Bechstedt}
 \affiliation{Institut f\"ur Festk\"orpertheorie und -optik, Friedrich-Schiller-Universit\"at, Max-Wien-Platz 1, D-07743 Jena, Germany}
\author{Marcelo Marques} 
 \email{gmsn@ita.br} \affiliation{Grupo de Materiais Semicondutores e Nanotecnologia, Instituto Tecnol\'ogico de Aeron\'autica, DCTA, 12228-900 S\~ao Jos\'e dos Campos, Brazil}
\author{Lara K. Teles}
 \email{gmsn@ita.br}
 \affiliation{Grupo de Materiais Semicondutores e Nanotecnologia, Instituto Tecnol\'ogico de Aeron\'autica, DCTA, 12228-900 S\~ao Jos\'e dos Campos, Brazil}

\date{\today}

\maketitle
\beginsupplement 

\section{Computational Details}

The structural and electronic properties of two-dimensional (2D) crystals and their heterostructures are calculated using the density functional theory (DFT) as implemented in the Vienna Ab-initio Simulation Package (VASP) \cite{Kresse1996a}.
The wave functions and pseudopotentials are generated within the projector-augmented wave (PAW) method \cite{Kresse1999}.
Exchange and correlation (XC) are described using the Perdew-Burke-Ernzerhof (PBE) functional within the generalized gradient approximation (GGA) \cite{Perdew1996}.
Van der Waals interaction is taken into account using the optB86b functional \cite{klime2011}.
The kinetic energy cutoff of the planewave expansion is restricted to 500~eV.
Integrations over the 2D Brillouin zone (BZ) are performed using an $11\times11\times1$ $\Gamma$-centered Monkhorst-Pack $\vec{k}$~point mesh \cite{Monkhorst1976} for $1\times1$ lateral unit cells.
The repeated slab method is applied to simulate individual 2D crystals as well as bilayer systems\cite{Bechstedt2003}.
A vacuum thickness of 15~\AA\ is employed to avoid unphysical interaction in the stacking direction.
Since a charge transfer may occur in vdW heterostructures, dipole corrections are applied to satisfy the periodic boundary conditions for the supercells.
Ionization energies \emph{I} and electron affinities \emph{A} of the isolated atomic layers are determined as differences of valence band maximum (VBM) and conduction band minimum (CBM), respectively to the vacuum level defined by vanishing electrostatic potential \cite{Koda2017}.
Natural band discontinuities $\Delta E_C$ ($\Delta E_V$) between the conduction band minima (valence band maxima) are determined as differences $A_A-A_B$ ($I_B-I_A$) in an AB heterostructure, where $A$ is the electronic affinity and $I$ is the ionization energy.

Minimum lateral unit cells employed are found using the coincidence lattice method \cite{Koda2016}.
Heterostructure investigations are performed after fixing the parameters of the most stable structural geometry for each monolayer and applying necessary strains to make the systems commensurate.
We make sure that the resulting biaxial strain in the 2D crystals is smaller than 2\% and that there are no more than 30 atoms inside the joint lateral cell.
All structural parameters are calculated first finding the energy minimum with a stopping criterion of $10^{-5}$~eV for the energy convergence and then relaxing the atomic positions until the Hellmann-Feynman forces on atoms are smaller than 1~meV/\AA.

Electronic properties calculated using the DFT functional lead to 2D band structures which suffer from the typical underestimation of energy gaps and interband distances computed as differences of Kohn-Sham eigenvalues of the DFT \cite{Paier2006}.
Therefore, they also lead to an incorrect description of hybridization and band offsets in vdW heterostructures \cite{Koda2017}.
To account for the excitation aspect we add approximate quasiparticle corrections to the Kohn-Sham bands by applying the XC hybrid functional HSE06 \cite{Paier2006,Heyd2003,Heyd2006}
to compute the electronic band structures and energy alignments.

\section{Influence of electric field}

\begin{figure}[b]
\centering
\includegraphics[width=0.5\textwidth]{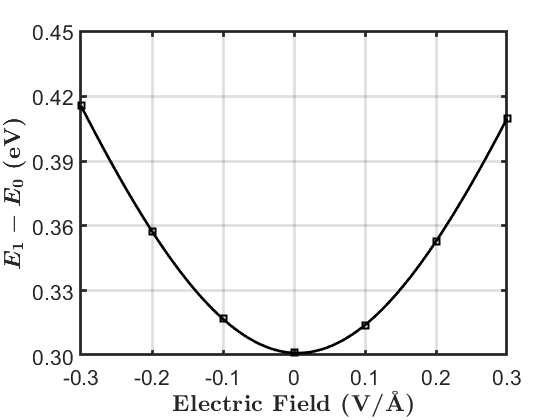}
\caption{Difference between the eigenenergies $E_1-E_0$ of the qubit as a function of the vertical electric field.}
\label{energysplit}
\end{figure}

\begin{figure*}
\centering
\includegraphics[width=\textwidth]{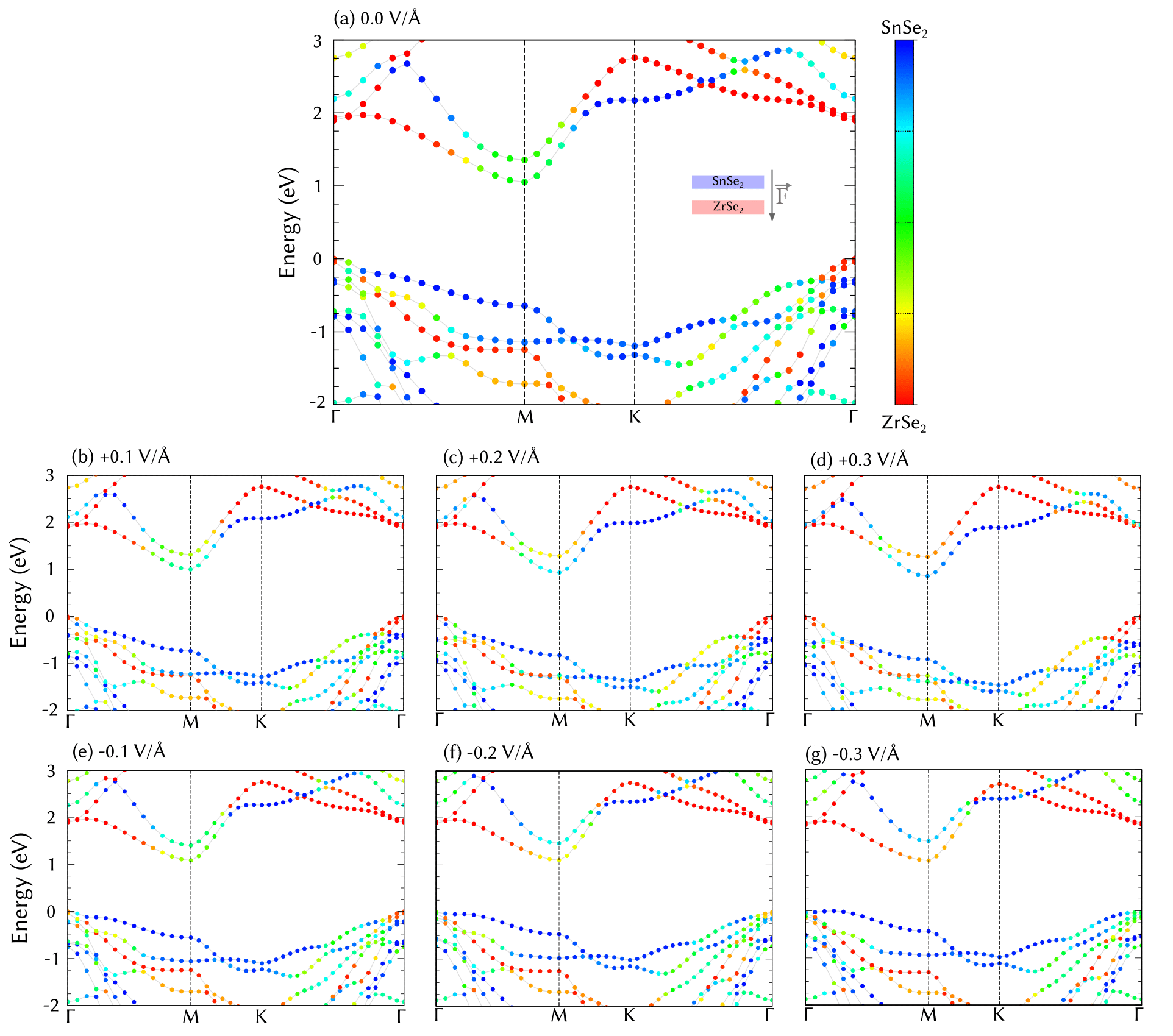}
\caption{Band structures of ZrSe$_2$/SnSe$_2$ heterostructure for different values of the vertical electric field. The inset in (a) shows the positive field orientation is considered from the ZrSe$_2$ layer to the SnSe$_2$ layer. The color of a dot in a Bloch band represents the relative contribution of each monolayer to the eigenvalue. The VBM is chosen as energy zero.}
\label{bands}
\end{figure*}

Our primary interest in this work is a two-level system formed by the states $\ket{0}$ and $\ket{1}$, which are derived from the two lowest conduction bands in a biased AB heterobilayer consisting of 2D crystals A and B.
We investigate a model heterostructure of transition metal dichalcogenide (TMDC) monolayers ZrSe$_2$ and SnSe$_2$.
One of the most important parameters of the heterosystem is the splitting of the two conduction band minima at M in the BZ.
The difference in energy between these two states as a function of the vertical electric field $F$ is plotted in Fig.~\ref{energysplit}.

The corresponding band structures for intermediate values of the electric field in Fig.~\ref{bands} allow us to identify trends in the shift of the eigenenergies, as well as in the change of the relative contributions of each layer to the eigenstates.
The presence of the electric field introduces just a small increase in the separation between the two levels, thereby drastically changing their character.
The actual field variation of $E_1-E_0$ in Fig.~\ref{energysplit} exhibits a nearly parabolic behavior, \emph{i.e.}, it is almost independent of the field orientation. Figure~\ref{bands} shows the evolution of the orbital character of all bands states around the fundamental gap for the model ZrSe$_2$/SnSe$_2$ bilayer system.


\begin{table*}
\centering
\caption{Difference in energy between the conduction and valence bands of several pairs of materials, \emph{i.e.}, the band discontinuities $\Delta E_C$ and $\Delta E_V$ (in eV). In green are the systems that present a misalignment lower than 0.3~eV, in yellow the ones that are between 0.3~eV and 0.6~eV and in red the ones that are greater than 0.6~eV. All the values are given in eV.}
\includegraphics[width=\textwidth]{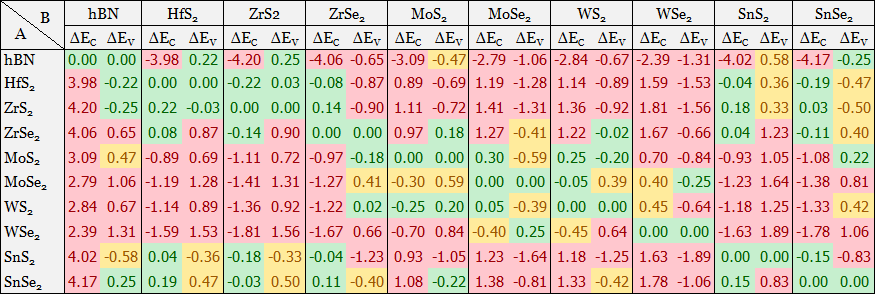}
\label{combinations}
\end{table*}

\begin{table*}
\centering
\caption{Difference in energy between the conduction band minimum of the materials in the columns with the valence band maximum of the materials in the rows. In green are the systems that present a misalignment lower than 0.3~eV, in yellow the ones that are between 0.3~eV and 0.6~eV and in red the ones that are greater than 0.6~eV. All the values are given in eV.}
\includegraphics[width=\textwidth]{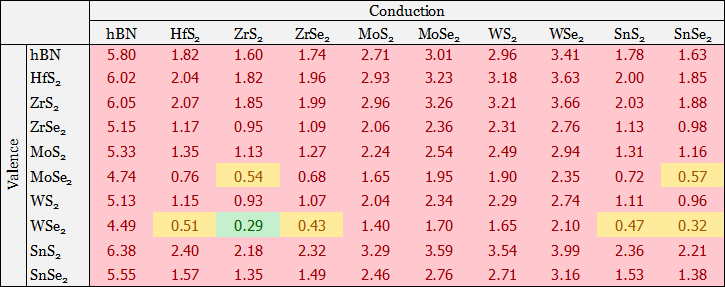}
\label{hybrid}
\end{table*}

\section{Search for other vdW qubits}

The weights $|\alpha_\psi(F)|^2$ and $|\beta_\psi(F)|^2$ are strongly related to the ``natural'' band discontinuities $\Delta E_C$ and $\Delta E_V$ of the band structures of the two sheet materials A and B forming the AB heterostructure, at least for vanishing gate field $F \rightarrow 0$ \cite{Koda2018}.
This has been demonstrated in Fig.~4 of the main text.
The Anderson's rule \cite{Anderson1962} is a first approximation to predict the alignment of the band structures of the individual sheets in the heterostructure.
Thus, it can be used as a first filter for selecting other possible AB heterosystems for charge vdW qubits.

The general result is that electron and hole distributions over the atomic sheets in a heterocombination can be only significantly modified by an external vertical electric field $F$ for small band discontinuities $\Delta E_C$ and $\Delta E_V$.
Considering field strengths of the order of $F=0.1\text{V/\AA}$ and vdW gaps between the sheets of about $d=3\text{\AA}$, field-induced modification of the band structure of energies $eFd=0.3$~eV appear as observed from Fig.~\ref{bands}.
Consequently, heterocombinations with maximum band discontinuities of about $|\Delta E_C|$ or $|\Delta E_V|\approx0.3$~eV may be switched with the gate voltage.

Besides the trivial match when the materials are the same, we observe that out of the 45 possible combinations, there are 13 with matching conduction bands and 9 with matching valence bands, as shown in Table \ref{combinations}.
Moreover, between these combinations there are only two, HfS$_2$/ZrS$_2$ and MoS$_2$/WS$_2$, with a simultaneous match between both the valence and conduction bands.
The explicit values are $\Delta E_C=-0.22~(0.25)$~eV and $\Delta E_V=0.03~(-0.20)$~eV for HfSe$_2$/ZrS$_2$~(MoS$_2$/WS$_2$), indicating that both heterostructures are of type II as also the model system ZrSe$_2$/SnSe$_2$.

Other extreme band alignments occur in the types II and III heterostructure cases, the staggered-gap and broken-gap systems, respectively, where the valence band of one material is aligned with the conduction band of the other material.
Among the 45 heterocombinations studied only the systems ZrSe$_2$/WSe$_2$ and SnSe$_2$/WSe$_2$ approach this situation, all of them being of type II.
The CBM of ZrS$_2$~(SnSe$_2$) is only 0.29~(0.32)~eV above the VBM of WSe$_2$, as shown in Table \ref{hybrid}.
Consequently, there should be a wave function overlap of the conduction-band functions of the A sheet and the valence-band functions of the B sheet.
A tunneling of electrons from the B=WSe$_2$ side into the A=ZrS$_2$ or SnSe$_2$ sheet should be possible.

Under the action of a negative gate field, this tendency will be enforced until strong tunneling of electrons from the valence band of one sheet into the conduction band of the other happens.
In the opposite field direction, the band alignment tends toward a pronounced type II heterostructure character.
Experimental studies of vdW heterostructure devices based on the WSe$_2$/SnSe$_2$ combination \cite{Roy2016} seem to imply that efficient carrier tunneling can be obtained by applying moderate gate voltages.
This is in line with the above theoretical predictions.

\begin{figure}
\centering
\includegraphics[width=0.5\textwidth]{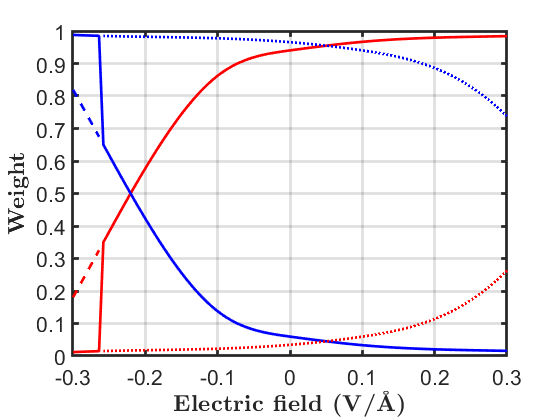}
\caption{Weights $|\alpha_h|^2$ (red, solid line) and $|\beta_h|^2$ (blue, solid line), as a function of the applied vertical electric field. The dashed lines indicate the weights for the maximum valence band state at $\Gamma$ and the dotted lines indicate the weights for the maximum valence band state at the reciprocal space position of the VBM for $F=-0.3$V/\AA.}
\label{hole}
\end{figure}

\section{Holes}

In order to work as a qubit as proposed, the ZrSe$_2$/SnSe$_2$ vdW heterostructure must have electrons in its conduction band.
One easy way to produce them is to excite electrons from the valence band via optical radiation, creating electron-hole pairs.
Therefore, in this case, the analysis of the charge on the sheets must include the effects of the hole wavefunction in the device operation.
The following analysis does not take into account excitonic effects, and thus give only general trends that might have small deviations from the actual behavior of the system.

As indicated in Fig.\ref{bands}, in the AB heterostructure, an electron in the lower CBM and a hole at the VBM move in the opposite direction, thereby, forming a vertical dipole.
This will not happen for an electron in the upper conduction state $\ket{1}$.

As the field becomes more negative, the bands of material B are shifted up.
At $F\approx-0.26$V/\AA, a bump in its valence band is raised above the energy level of the maximum valence band state at $\Gamma$, resulting in the VBM of the heterostructure being shifted from $\Gamma$ to a point between $\Gamma$ and M, as indicated in Fig.~\ref{bands}g.
This causes an abrupt change in the weights $\alpha_h$ and $\beta_h$, since the characters of these states are substantially different for this value of $F$, as indicated in Fig.~\ref{hole}.

The fact that the holes change their localization from one sheet to the other depending on the electric field value makes the comparison between the electrical current of electrons not straightforward, so engineering the (electro)chemical potential position by n-type doping would be preferable than producing free carriers by optical absorption transitions.


\bibliography{references}